\documentclass[12pt]{article}
\usepackage{amsmath,amsthm,latexsym,amssymb,multibox}
\newcommand{\be}{\begin{equation}}
\newcommand{\ee}{\end{equation}}
\newcommand{\ba}{\begin{eqnarray}}
\newcommand{\ea}{\end{eqnarray}}
\newcommand{\pr}{\partial }

\def\nn{\nonumber}
\def\q{\quad}

\def\d{\delta}

\def\h{\eta}

\def\f{\phi}

\def\m{\mu}
\def\n{\nu}
\def\r{\rho}

\def\IR{\relax{\rm I\kern-.18em R}}
\def\ZZ{\relax{\hbox{\cmss Z\kern-.4em Z}}}

\begin{document}

\begin{titlepage}
\begin{flushright}
DFPD/03/TH/06\\
ULB--TH--03/05\\
hep-th/0301243\\
\end{flushright}

\vspace{2cm}

\begin{center} {\Large{\bf On geometric equations and duality \\ for free higher spins}}
\end{center}
\vspace{1.3cm}
\begin{center} {\large Xavier Bekaert$^\clubsuit$ and Nicolas Boulanger$^\star$\footnote{``Chercheur
F.R.I.A.'', Belgium} }
\end{center}

\begin{center}{\sl
$^\clubsuit$ Dipartimento di Fisica, Universit\`a degli Studi di Padova\\ Via
F. Marzolo 8, 35131 Padova, Italy}\end{center}
\begin{center}{\sl
$^\star$ Facult\'e des Sciences, Universit\'e Libre de Bruxelles,\\
Campus Plaine C.P. 231, B--1050 Bruxelles, Belgium }\end{center}

\vspace{.3cm}

\begin{abstract}
We provide a general scheme for dualizing higher-spin gauge fields
in arbitrary irreducible representations of $GL(D,\mathbb R)$. We
also give a recipe for constructing Fronsdal-like field equations
and Lagrangians for such exotic fields.
\end{abstract}

\vfill
\end{titlepage}
\section{Introduction}

Despite several decades of study, higher-spin gauge fields (i.e.
spin $S>2$) are still rather mysterious. For instance,
weakly-coupled ${\cal N}=4$ super-Yang-Mills was recently
conjectured to be the holographic dual of a theory with infinitely
many higher-spin fields in $AdS_5$ \cite{Mikhailov:2002}. In any
case, the old Fronsdal program \cite{Fronsdal:1978} of
constructing interactions of massless higher-spin fields by
introducing consistent coupling with sources is far from being
achieved. It was formulated in the late seventies when Fang and
Fronsdal obtained the covariant Lagrangians for gauge fields of
any spin in a flat background
\cite{Fronsdal:1978,Fronsdal:1978bis}. It was soon followed by an
alternative approach to free massless higher-spin fields, the so
called ``gauge approach" introduced by Vasilev
\cite{Vasiliev:1980}, which uses geometrical objects generalizing
vielbein and spin connection. This approach turned out to be
promising for switching on consistent interactions
\cite{Fradkin:1986}. In a recent work on higher-spin gauge fields,
Francia and Sagnotti \cite{Francia:2002,Francia:2002bis}
discovered that forgoing locality allows to relax the trace
conditions of the Fang-Fronsdal formulation. For arbitrary spin
$S$, gauge invariant field equations were elegantly written in
terms of the curvature tensor introduced by de Wit and Freedman
\cite{deWit}.

In four dimensions, all the tensorial irreps of the little group
$SO(2)$ are completely symmetric and the rank $S$ is equal to the
spin. In dimension $D>4$, other irreps are possible and, in such
cases, the ``spin'' $S$ loosely refers to the number $S$ of
columns in the corresponding Young diagram. Tensor fields in
arbitrary irreps of the Lorentz group appear in the spectrum of
string theory. One may also dualise in the light-cone gauge some
of the physical components of a completely symmetric tensor, which
naturally leads to ``exotic'' irreps of the little group
$SO(D-2)$. These duality transformations can be performed
covariantly by acting with the (space-time) Levi-Civita tensor on
the curvature tensor, exchanging thereby the role played by
Bianchi identities and field equations \cite{Hull:2001}. Guided by
the duality symmetry principle, a systematic study led to
conjectured field equations for tensor gauge field theories in
arbitrary irreps of $GL(D,\mathbb R)$ \cite{Bekaert:2002}. An
important object was introduced which generalizes de
Wit-Freedman's curvature for tensor gauge fields in arbitrary
irreps of $GL(D,\mathbb R)$. In \cite{deMedeiros:2002}, de
Medeiros and Hull followed another path: they constructed field
equations for exotic fields by deriving them from gauge invariant
Lagrangians. Doing so, they obtained a higher-derivative version
of Francia-Sagnotti's field equations for any irrep. of
$GL(D,\mathbb R)$. Retrospectively, this part of their work can be
seen as the generalization of the work of Francia and Sagnotti for
exotic gauge fields.

All known formulations of free massless higher spin fields exhibit
new features with respect to spin $S\leq 2$ fields (e.g. trace
conditions, non-locality, higher derivative kinetic operators,
auxiliary fields, etc). These unavoidable novelties of higher spins
are deeply rooted in the fact that the curvature tensor, that might
be the central object in higher-spin theory, contains $S$
derivatives. A major progress of the recent approaches was to
produce ``geometric" field equations, i.e. equations written
explicitly in terms of the curvature.

The main result of this paper is to provide an explicit
relationship between (1) a duality-symmetric approach to free
higher spins fields and (2) the old local approach of Fronsdal
(reviewed in section \ref{Fron}), as well as (3) the recent
non-local approach of Francia and Sagnotti (reviewed in section
\ref{Sagn}). The equivalence of (1) with (2) and (3) is shown in
section \ref{dualsym}. As an important by-product of this result,
we obtain a covariant method for dualising free higher-spin
fields. In section \ref{exotic}, we analyze in detail an exotic
tensor gauge theory which is dual to standard spin-three gauge
theory in five dimensions. For definiteness, we indeed concentrate
on the spin-three field and comment on the general case at the
end, section \ref{last}.

\section{Local approach}\label{Fron}

The main advantage of the Fang-Fronsdal approach to free massless
fields is that it respects two requirements of orthodox quantum
field theory:
\begin{itemize}
  \item (i) Locality, and
  \item (ii) Second order field equations (for bosonic fields).
\end{itemize}
The spin-three Fronsdal equation \cite{Fronsdal:1978} reads \be
{\cal F}_{\m_1\m_2\m_3} \equiv \Box \phi_{\m_1\m_2\m_3} -
3\pr_{(\m_1} \pr^{\m_4}\phi_{\m_2\m_3)\m_4} +3
\pr_{(\m_1}\pr^{}_{\m_2}\phi_{\m_3)\m_4}^{\quad\quad\m_4} =0\,
\label{frspin3}\ee where the parenthesis $(\,\,)$ means the
symmetrization with weight one. Indices are raised and lowered with
the Minkowski metric $\eta_{\m_1\m_2}$. The spin-3 field
$\phi_{\m_1\m_2\m_3}$ is completely symmetric:
$\phi_{(\m_1\m_2\m_3)}=\phi_{\m_1\m_2\m_3}$. Its gauge
transformations are of the form\be \delta \phi_{\m_1\m_2\m_3} \
= \ 3\partial_{(\m_1} \Lambda_{\m_2\m_3)} \ . \label{deltaphi3}
\ee
But since (\ref{deltaphi3}) transforms ${\cal F}$ as
\be
\delta {\cal F}_{\m_1\m_2\m_3} \ = \ 3 \,
\partial_{\m_1}
\partial_{\m_2} \partial_{\m_3} \,
\Lambda_{\quad\m_4}^{\m_4} \,,
 \label{deltaF}
\ee
the
gauge parameter $\Lambda_{\m_1\m_2}$ is constrained to be
traceless in order to leave the field equation (\ref{frspin3})
invariant. Eventually, the standard de Donder gauge-fixing
condition \be \pr^{\m_4}\phi_{\m_2\m_3\m_4} -
\pr^{}_{(\m_2}\phi_{\m_3)\m_4}^{\quad\quad\m_4} =0\ee is used to
reduce the Fronsdal equation to its canonical form $\Box
\phi_{\m_1\m_2\m_3}=0$. As shown in \cite{deWit}, this gauge
theory leads to the correct number of physical degrees of freedom:
$\frac16(D+2)(D-2)(D-3)$, that is the dimension of the irrep. of
the little group $SO(D-2)$ corresponding to the Young diagram
$(1,1,1)$.

\section{Non-local approach}\label{Sagn}

The trace condition on the gauge parameter looks simple (even if
somewhat unnatural) but, unfortunately, it proved to be
technically involved to deal with. Recently, gauge invariant field
equations for unconstrained parameters were constructed by Francia
and Sagnotti, at the price of loosing locality. For spin three, a
simple form of their field equation is the following
\cite{Francia:2002}\be {\cal F}_{\m_1\m_2\m_3} \, = \,
\partial_{\m_1}\partial_{\m_2} \partial_{\m_3} {\cal H}\label{FS1}\ee where $\cal H$ is given by
the non-local expression \be {\cal
H}\,\equiv\,\frac{1}{\Box^2}\,\partial^{\m_1}{\cal
F}_{\m_1\m_2}^{\quad\quad\m_2}\,.\ee As one can see, the
requirement (i) of section \ref{Fron} is left over, however the
requirement (ii) is still satisfied. The gauge invariance of
(\ref{FS1}) under unconstrained parameter is easily checked since
(\ref{deltaF}) implies that \be \delta{\cal H}=3
\Lambda_{\,\,\,\m}^{\m}\,.\ee Of course, Fronsdal equation is
easily recovered by setting ${\cal H}=0$ by an appropriate gauge
transformation. This choice restores locality and fixes the trace
of the gauge parameter, as it should be in the Fronsdal approach.
In conclusion, the Fronsdal approach is obtained as a gauge-fixing
of the Francia-Sagnotti formulation.


As noticed in \cite{Francia:2002}, it turns out that the equation
(\ref{FS1}), once combined with its trace, can elegantly be
expressed in terms of the curvature tensor introduced by de Wit
and Freedman \cite{deWit} (the precise expression is given at the
end of the section). The ``curvature'' is meant for a
gauge-invariant object constructed from the field, the vanishing
of which implies that the field is pure gauge, i.e.
$\phi_{\m_1\m_2\m_3}=\partial_{(\m_1} \Sigma_{\m_2\m_3)}$. With
duality in mind, we consider the most natural gauge-invariant
object under (\ref{deltaphi3}), which is obtained by taking three
curls, i.e. one curl for every index of the field\be {\cal
K}_{\m_1\n_1\,\mid\,\m_2\n_2\,\mid\,\m_3\n_3}=
8\,\partial_{[\m_1}\phi_{\n_1]\,\mid\,[\m_2,\n_2]\,\mid\,[\m_3,\n_3]}\,.\label{curvature}\ee
The bracket $[\,\,]$ stands for the antisymmetrisation (with
weight one) over a pair of indices $(\m_i,\n_i)$. By construction,
this tensor is antisymmetric under the exchange of two indices in
any given pair $(\m_i,\n_i)$ \be {\cal
K}_{[\m_1\n_1]\,\mid\,\m_2\n_2\,\mid\,\m_3\n_3}={\cal
K}_{\m_1\n_1\,\mid\,[\m_2\n_2]\,\mid\,\m_3\n_3}={\cal
K}_{\m_1\n_1\,\mid\,\m_2\n_2\,\mid\,[\m_3\n_3]}={\cal
K}_{\m_1\n_1\,\mid\,\m_2\n_2\,\mid\,\m_3\n_3}\,.\ee Furthermore,
it is obviously symmetric under the exchange of two pairs \be
{\cal K}_{\m_1\n_1\,\mid\,\m_2\n_2\,\mid\,\m_3\n_3}={\cal
K}_{\m_2\n_2\,\mid\,\m_1\n_1\,\mid\,\m_3\n_3}={\cal
K}_{\m_3\n_3\,\mid\,\m_2\n_2\,\mid\,\m_1\n_1}={\cal
K}_{\m_1\n_1\,\mid\,\m_3\n_3\,\mid\,\m_2\n_2}\,.\ee In other words
the curvature tensor belongs to the irrep. of $GL(D,\mathbb R)$
corresponding to the Young diagram $(2,2,2)$. The curvature tensor
also satisfies two types of cyclic identities: algebraic ones
(called ``first Bianchi identities'') where we antisymmetrize any
three indices \be {\cal
K}_{[\m_1\n_1\,\mid\,\m_2]\n_2\,\mid\,\m_3\n_3}=0\,, \label{BI}
\ee and differential ones (christened as ``second Bianchi
identities'') where one takes a curl over any pair of indices \be
\pr_{[\r_1}{\cal
K}_{\m_1\n_1]\,\mid\,\m_2\n_2\,\mid\,\m_3\n_3}=0\,.\label{BII}\ee
These properties directly generalize the well-known properties of
the linearized Riemann tensor.

In the very inspiring work \cite{Damour}, Damour and Deser proved
that the vanishing of curvature (\ref{curvature}) indeed implies
that the spin-three field is pure gauge. Moreover, they showed
that, if in addition Fronsdal equation (\ref{frspin3}) is
satisfied, then the gauge parameter can always be taken to be
traceless\footnote{We stress that analogous ``generalized
Poincar\'e lemmas'' can be shown to hold for rank $S$ symmetric
tensors by using the results of \cite{Dubois-Violette:1999}.
First, the vanishing of the curvature implies that the field is
pure gauge
$\phi_{\m_1\m_2\ldots\m_S}=\pr_{(\m_1}^{}\Lambda_{\m_2\ldots\m_S)}$
with an unconstrained gauge parameter
$\Lambda_{\m_1\ldots\m_{S-1}}$. Second, the Fronsdal equation
${\cal
F}_{\m_1\m_2\ldots\m_S}=\pr_{(\m_1}^{}\pr_{\m_2}^{}\pr_{\m_3}^{}
\Lambda_{\m_4\ldots\m_S)\m_{S+1}}^{\quad\quad\quad\quad\quad\m_{S+1}}=0$
implies that the trace
$\Lambda_{\m_1\ldots\m_{S-3}\m_{S-2}}^{\quad\quad\quad\quad\quad\m_{S-2}}$
of the gauge parameter is a polynomial of order $< S$.
Unfortunately, one may assume without loss of generality that the
latter vanishes only in the very specific spin-three case.}. This
last result was one of the first direct manifestation of the
curvature tensor relevance in higher-spin gauge theory. Today one
can argue that the curvature already plays a significant role in
the Fronsdal approach but is somehow ``hidden'', due to the
conjugated requirements (i) and (ii) of section \ref{Fron}.

The curvature tensor ${\cal R}_{\m_1\n_1\r_1\,;\,\m_2\n_2\r_2}$ of
de Wit and Freedman and the ``Riemann tensor'' ${\cal
K}_{\m_1\n_1\,\mid\,\m_2\n_2\,\mid\,\m_3\n_3}$ of Damour and Deser
are related by acting with the appropriate Young symmetrizers \be
{\cal R}^{\m_1\n_1\r_1}_{\quad\quad ;\,\,\m_2\n_2\r_2}\equiv {\cal
K}^{\m_1\quad\quad\n_1\quad\r_1}_{\quad(\m_2\,\mid\quad\n_2\,\mid\quad\r_2)}\,,\label{dWF}\ee
and \be {\cal
K}_{\m_1\n_1\,\mid\,\m_2\n_2\,\mid\,\m_3\n_3}=2\,{\cal
R}_{[\m_1[\m_2[\m_3\,;\,\,\n_1]\n_2]\n_3]}\,,\ee where the three
antisymmetrizations are taken over every pair of indices
$(\m_i,\n_i)$. The de Wit-Freedman tensor is, by construction,
symmetric in each of the two sets of three indices \be {\cal
R}_{(\m_1\n_1\r_1)\,;\,\,\m_2\n_2\r_2}={\cal
R}_{\m_1\n_1\r_1\,;\,\,(\m_2\n_2\r_2)}={\cal
R}_{\m_1\n_1\r_1\,;\,\,\m_2\n_2\r_2}\,.\ee Finally, we mention
that (\ref{FS1}) is equivalent to the geometric equation
\cite{Francia:2002} \be {\cal F}_{\m_1\m_2\m_3} -
\frac{1}{\Box}\,\pr^{}_{(\m_1}\pr^{}_{\m_2}{\cal
F}_{\m_3)\m_4}^{\quad\quad\m_4}\,\equiv\,
\frac{1}{\Box}\,\,\pr^{\n_1}{\cal
R}^{\n_2}_{\,\,\,\,\,\n_1\n_2\,;\,\m_1\m_2\m_3} =0\,.
\label{FSbis}
\ee

\section{Duality-symmetric approach}\label{dualsym}

For duality purposes, the curvature tensor ${\cal
K}_{\m_1\n_1\,\mid\,\m_2\n_2\,\mid\,\m_3\n_3}$ is the natural
object to consider since it displays the appropriate symmetries.
Indeed, one can dualise on every pair of antisymmetric indices.

A decisive step is now to express spin-three field equations as
Einstein-like equations in order to follow the scheme developed in
\cite{Hull:2001,Bekaert:2002}. More precisely, the idea was to
generalize the approach of \cite{Hull:2001} for spin-two where
linearized Einstein equations were considered as Bianchi
identities for the dual theory. On the other hand a deep
relationship between ${\cal F}_{\m_1\m_2\m_3}$ and the trace of
the curvature ${\cal
K}_{\m_1\n_1\,\mid\,\m_2\n_2\,\mid\,\m_3\n_3}$ was discovered by
Damour and Deser \be \pr_{[\m_1}{\cal
F}_{\n_1]\m_2\m_3}=\eta^{\n_2\n_3}{\cal
K}_{\m_1\n_1\,\mid\,\m_2\n_2\,\mid\,\m_3\n_3}\,.\label{DD}\ee
Therefore, the ``Einstein" equation \be {\cal
K}^{\quad\n_1}_{\m_1\quad\,\mid\,\m_2\n_1\,\mid\,\m_3\n_3}=0\,
\label{Einstein} \ee is a consequence of Fronsdal equation
(\ref{frspin3}) as well as Francia-Sagnotti equation (\ref{FS1}).
Conversely, the field equation (\ref{Einstein}) states that any
curl of ${\cal F}_{\m_1\m_2\m_3}$ vanishes but, according to the
analysis of \cite{Bekaert:2002,Dubois-Violette:1999}, this is
equivalent to \be {\cal F}_{\m_1\m_2\m_3} \ = \ \partial_{\m_1}
\partial_{\m_2} \partial_{\m_3} \,
\Sigma\,. \ee Due to (\ref{deltaF}), it is clear that one recovers
Fronsdal's approach by performing a gauge transformation with
parameter $\Lambda_{\m\n}$, the trace of which is fixed to be
$3\Lambda_{\quad\m}^{\m}=-\Sigma$. On the other hand, from
(\ref{dWF}) one immediately sees that the field equation
(\ref{Einstein}) implies the vanishing of the trace of de
Wit-Freedman's tensor, thereby ensuring Francia-Sagnotti field
equation (\ref{FSbis}). To summarize, the field equation
(\ref{Einstein}) is an equivalent form of the field equation for
the spin-three field. Its drawback is that it does not immediately
derive from an action principle due to an inappropriate number of
free indices. But from the point of view of duality, it is the
most suitable form since it can be read as the following first
Bianchi identity\be (*{\cal
K})_{[\m_1\ldots\m_{D-2}\,\mid\,\m_{D-1}]\n\,\mid\,\r_1\r_2}=0\label{BIdual}
\ee for the dual curvature tensor \be (*{\cal
K})_{\m_1\ldots\m_{D-2}\,\mid\,\n_1\n_2\,\mid\,\r_1\r_2}\equiv
\frac{1}{2}\,\epsilon_{\m_1\ldots\m_{D-2}\m_{D-1}\m_D}\,{\cal
K}^{\m_{D-1}\m_D}_{\quad\quad\quad\mid\,\n_1\n_2\,\mid\,\r_1\r_2}\,,\ee
where $\epsilon_{\m_1\ldots\m_D}$ is the Levi-Civita tensor.
Another first Bianchi identity for the dual curvature follows
directly from the first Bianchi identity (\ref{BI}) for the
original curvature\be (*{\cal
K})_{\m_1\ldots\m_{D-2}\,\mid\,\n[\r_1\,\mid\,\r_2\r_3]}=0\,.\label{BIdual'}\ee
Together, these two dual Bianchi identities (\ref{BIdual}) and
(\ref{BIdual'}) imply that the dual curvature tensor is in the
irrep. of $GL(D,\mathbb R)$ corresponding to the diagram
$(D-2,2,2)$. Of course, in four dimensions, the dual tensor has
the same symmetry properties than the original curvature tensor.

Let us now rewrite the second Bianchi (\ref{BII}) in terms of the
dual curvature tensor \be\pr^{\m_1}(*{\cal
K})_{\m_1\m_2\ldots\m_{D-2}\,\mid\,\n_1\n_2\,\mid\,\r_1\r_2}=0\ee
Due to the cyclic property (\ref{BIdual}) we get that the
divergence of the curvature tensor vanishes on-shell
\be
\pr^{\m_1}{\cal
K}_{\m_1\n_1\,\mid\,\m_2\n_2\,\mid\,\m_3\n_3}=0\,.
\ee
In terms of
the dual curvature tensor this translates into the second Bianchi
identity
\be
\pr_{[\m_1}(*{\cal
K})_{\m_2\ldots\m_{D-1}]\,\mid\,\n_1\n_2\,\mid\,\r_1\r_2}=0\,.
\label{BIIdual}
\ee
Another second Bianchi identity
\be (*{\cal
K})_{\m_1\ldots\m_{D-2}\,\mid\,\n_1\n_2\,\mid\,[\r_1\r_2,\r_3]}=0\,.
\label{BIIdual'}
\ee
directly follows from (\ref{BII}). The second Bianchi identities
(\ref{BIIdual}) and (\ref{BIIdual'}) together with the (on-shell)
symmetry property of the dual curvature tensor imply that
\cite{Bekaert:2002}
\be
(*{\cal
K})_{\m_1\ldots\m_{D-2}\,\mid\,\n_1\n_2\,\mid\,\r_1\r_2}=
\partial_{[\m_1}\tilde{\phi}_{\m_2\ldots\m_{D-2}]\,\mid\,[\n_1,\n_2]
\,\mid\,[\r_1,\r_2]}\,
\ee
for a dual gauge field
$\tilde{\phi}_{\m_1\ldots\m_{D-3}\,\mid\,\n\,\mid\,\r}$ with the
symmetry properties corresponding to the Young diagram $(D-3,1,1)$
that is, which satisfies
\be
\tilde{\phi}_{[\m_1\ldots\m_{D-3}]\,\mid\,\n\,\mid\,\r}=\tilde{\phi}_{\m_1\ldots\m_{D-3}\,\mid\,(\n\,\mid\,\r)}
=\tilde{\phi}_{\m_1\ldots\m_{D-3}\,\mid\,\n\,\mid\,\r}\,,\quad
\tilde{\phi}_{[\m_1\ldots\m_{D-3}\,\mid\,\m_{D-2}]\,\mid\,\n}=0\,.
\ee
In other words, the tensor $(*{\cal
K})_{\m_1\ldots\m_{D-2}\,\mid\,\n_1\n_2\,\mid\,\r_1\r_2}$ is
indeed a curvature for the dual gauge field
$\tilde{\phi}_{\m_1\ldots\m_{D-3}\,\mid\,\n\,\mid\,\r}$.
Finally, the two field equations of the dual field theory are
\ba\eta^{\m_1\n_1}(*{\cal
K})_{\m_1\ldots\m_{D-2}\,\mid\,\n_1\n_2\,\mid\,\r_1\r_2}&=&0\,,\label{dualeom1}\\
\eta^{\n_1\r_1}(*{\cal
K})_{\m_1\ldots\m_{D-2}\,\mid\,\n_1\n_2\,\mid\,\r_1\r_2}&=&0\,.\label{dualeom2}\ea
The equations (\ref{dualeom1}) and (\ref{dualeom2}) are
respectively obtained from (\ref{BI}) and (\ref{Einstein}).

In four dimensions, we recover a usual spin-three field
$\tilde{\phi}_{\m_1\m_2\m_3}$. In five dimensions, the dual theory
is an exotic spin-three theory with gauge field
$\tilde{\phi}_{\m_1\m_2\,\mid\,\n\,\mid\,\r}$. This theory is
investigated in detail in the next section. To end up the present
section, we mention the two other possibilities: one could have
either dualised the curvature over two pairs of antisymmetric
indices getting a dual gauge field belonging to the irreducible
representation $(D-3,D-3,1)$ of $GL(D,\mathbb R)$, or one could
have dualised all pairs leading to a $(D-3,D-3,D-3)$ tensor gauge
field theory \cite{Hull:2001,Bekaert:2002}.

\section{Exotic spin-three gauge theory}\label{exotic}

The curvature for a gauge field
$\phi_{\m_1\n\,\mid\,\m_2\,\mid\,\m_3}$ with symmetries $(2,1,1)$
is given by \be {\cal
K}_{\m_1\n_1\r\,\mid\,\m_2\n_2\,\mid\,\m_3\n_3}\equiv
\partial_{[\m_1}\phi_{\n_1\r]\,\mid\,[\m_2,\n_2]\,\mid\,[\m_3,\n_3]} \label{K}
\ee and possesses the symmetries $(3,2,2)$. Generalizing the
analysis of Damour and Deser, the crucial step is to {\it{define}}
a Fronsdal tensor ${\cal F}_{\m_1\n\,\mid\,\m_2\,\mid\,\m_3}$ in
the irrep. $(2,1,1)$ as follows
\be
{\cal
F}_{\m_1\n_1\,\mid\,\m_2\,\mid\, [\m_3,\n_3]}\equiv
\h^{\r\n_2}{\cal
K}_{\m_1\n_1\r\,\mid\,\m_2\n_2\,\mid\,\m_3\n_3}\,.
\label{definericci}
\ee
The result is
\ba {\cal F}_{\m_1\n\,\mid\,\m_2\,\mid\,\m_3}&=& \Box
\f_{\m_1\n\,\mid\,\m_2\,\mid\,\m_3}
-2\pr^{\m_4}\f_{\m_1\n\,\mid\,\m_4\,\mid\,(\m_2,\m_3)} +\pr_{\m_2}
\pr_{\m_3} \f_{\m_1\n\,\mid\,\m_4\,\mid\,}^{\q\q\q\m_4}
\nonumber \\
&&+2\pr^{\r}\pr_{[\m_1}\f_{\n]\r\,\mid\,\m_2\,\mid\,\m_2}
-4\pr^{}_{[\m_1}\f_{\n]\r\,\mid\,\,\,\,\mid\,(\m_2,\m_3)}^{\,\,\q\r}\,.
\label{ricci} \ea We take \be
\h^{\m_1\m_2}K_{\m_1\n_1\r\,\mid\,\m_2\n_2\,\mid\,\m_3\n_3}=0
\label{eom} \ee as our equation of motion in dimension $D\geq 5$.
Due to the symmetry properties of the curvature, the other two
traces also vanish. The equations (\ref{eom}) and
(\ref{definericci}) imply that the Fronsdal tensor must be written
as \be {\cal F}_{\m_1\n\,\mid\,\m_2\,\mid\,\m_3}=
\pr_{\m_2}\pr_{\m_3}C_{[\m_1\n]}\,. \ee Now, the only way to match
the symmetries of both sides of the above equation is through \be
{\cal F}_{\m_1\n\,\mid\,\m_2\,\mid\,\m_3}=
\pr_{\m_2}\pr_{\m_3}\pr_{[\m_1}C_{\n]}\,. \label{truc} \ee Another
way to obtain this identity is by noticing that setting the curl
of a $GL(D,\mathbb R)$-irreducible tensor (i.e. the Fronsdal
tensor) to zero, ${\cal
F}_{\m_1\n\,\mid\,\m_2\,\mid\,[\m_3,\r]}=0$, where the curl is
taken on the last column of the tensor, is equivalent to imposing
that all the {\it{irreducible}} components with 3 columns of the
first derivative of the tensor must be set to zero. This type of
cocycle conditions for ${\cal F}_{\m_1\n\,\mid\,\m_2\,\mid\,\m_3}$
directly implies (\ref{truc}), using the results of
\cite{Bekaert:2002}.
\\
The gauge transformations associated to the curvature are
\cite{Bekaert:2002}\be \d
\f_{\m_1\n\,\mid\,\m_2\,\mid\,\m_3}=3\pr_{[\m_1}S_{\n]\m_2\m_3}
+\frac{3}{2}M_{\m_1\n\,\mid\,(\m_2,\m_3)}-\pr_{[\m_1}M_{\n](\m_2\,\mid\,\m_3)}
\label{tj} \ee where $S_{\m_1\m_2\m_3}$ is completely symmetric
and $M_{\m_1\n\,\mid\,\m_2}$ has the mixed symmetry $(2,1)$. The
reducibilities are \be \d \f_{\m_1\n\,\mid\,\m_2\,\mid\,\m_3}=0
\q\Leftrightarrow\q S_{\m_1\m_2\m_3}=\pr_{(\m_1}S_{\m_2\m_3)}~;~~
M_{\m_1\n\,\mid\,\m_2}=-\pr_{[\m_1}S_{\n]\m_2} \label{redu} \ee
for the symmetric reducibility parameter $S_{\m_1\m_2}$. Then the
gauge variation of the Fronsdal tensor is \be \d {\cal
F}_{\m_1\n\,\mid\,\m_2\,\mid\,\m_3}=
\pr_{\m_2}\pr_{\m_3}\pr_{[\m_1}\Big(
3S_{\n]\r}^{\q\r}-4M_{\n]\r\,\mid\,}^{\q\,\r}\Big). \ee This,
together with (\ref{truc}), shows that an appropriate gauge choice
enables us to reach the Fronsdal gauge \be {\cal
F}_{\m_1\n\,\mid\,\m_2\,\mid\,\m_3}= 0\,. \label{fieldequ} \ee The
Fronsdal tensor (\ref{ricci}) is gauge-invariant for \be
3S_{\n\r}^{\q\r}=4M_{\n\r\,\mid\,}^{\q\,\r}\label{gfSM}\ee which
relates the traces of the two gauge parameters. This constraint is
preserved by the ``reducibility transformations'' (\ref{redu}) if
and only if \be \h^{\m_1\m_2}S_{\m_1\m_2}=0\,. \label{consist} \ee
With the covariant de Donder condition \be
D_{\m_1(\m_2\m_3)}\equiv
\f_{\m_1\n\,\mid\,\,\,\,\mid\,(\m_2,\m_3)}^{\q\,\,\,\,\,\n}-\pr^{\n}\f_{\m_1\n\,\mid\,\m_2\,\mid\,\m_3}=0
\label{dedon} \ee which contains as many degrees of freedom as the
ones contained in $S_{\m\n\r}$ and $M_{\m\n\,\mid\,\r}$, the field
equation takes its canonical form \be \Box
\f_{\m_1\n\,\mid\,\m_2\,\mid\,\m_3}= 0\,. \ee The gauge variation
of the de Donder condition (\ref{dedon}) gives \be \d
D_{\m_1(\m_2\m_3)}=\frac12\Box(3 S_{\m_1\m_2\m_3}-
M_{\m_1(\m_2\,\mid\,\m_3)}) \ee provided one imposes the following
covariant de Donder condition for the gauge parameters \be
D_{(\m_1\m_2)}\equiv
\pr^{\nu}\Big(S_{\m_1\m_2\nu}-\frac{1}{3}\,M_{\nu(\m_1\,\mid\,\m_2)}\Big)
-\frac{3}{4}\,\pr_{(\m_1}S_{\m_2)\nu}^{\q\,\,\,\nu}=0\,.
\label{dedon2} \ee which is traceless
$\eta^{\m_1\m_2}D_{\m_1\m_1}=0$ when (\ref{gfSM}) is satisfied.
The variation of $D_{(\m_1\m_2)}$ is given by \be \d
D_{(\m_1\m_2)}=\frac{3}{4}\Box S_{\m_1\m_2}\,,\ee where one used
the consistency condition (\ref{consist}). We then obtained the
appropriate gauge conditions which give
$\frac{1}{8}(D-4)(D-3)(D-1)(D+2)$ for the number of physical
degrees of freedom \cite{Labastida:1987}. In five dimensions, we
indeed have the $7$ physical degrees of freedom of the standard
spin-three field.

The Fronsdal-like equation (\ref{fieldequ}) can be derived from
the action
\be
S[\f_{\m_1\n\,\mid\,\m_2\,\mid\,\m_3}]=\frac12\int
\f_{\m_1\n\,\mid\,\m_2\,\mid\,\m_3} {\cal
G}^{\m_1\n\,\mid\,\m_2\,\mid\,\m_3} \label{action} \,,
\ee
where we
introduced the ``Einstein" tensor
\ba
{\cal G}_{\m_1\n\,\mid\,\m_2\,\mid\,\m_3}&\equiv&{\cal
F}_{\m_1\n\,\mid\,\m_2\,\mid\,\m_3}-\frac{1}{2} \,({\cal
F}^1_{\m_1\n}\h_{\m_2\m_3}+{\cal
F}^1_{[\m_1\,\mid\,\m_2}\h_{\n]\m_3}+{\cal
F}^1_{[\m_1\,\mid\,\m_3}\h_{\n]\m_2})\nn\\&& -\h_{\m_2[\m_1}{\cal
F}^2_{\n]\m_3}-\h_{\m_3[\m_1}{\cal F}^2_{\n]\m_2}\,, \ea with the
two linearly independent traces ${\cal F}^1_{\m_1\m_2}\equiv
\h^{\m_3\m_4}{\cal F}_{\m_1\m_2\,\mid\,\m_3\,\mid\,\m_4}$ and
${\cal F}^2_{\m_1\m_2}\equiv \h^{\m_3\m_4}{\cal
F}_{\m_3(\m_1\,\mid\,\m_2)\,\mid\,\m_4}$. The Einstein tensor
possesses the symmetries of the field, defines a self-adjoint
(second-order) differential operator in (\ref{action}) and satisfies
\be
\pr^{\m_1}{\cal G}_{\m_1\n\,\mid\,\m_2\,\mid\,\m_3}
=-\frac{1}{2}\,\h_{\m_2\m_3}\pr^{\m_1}{\cal F}^1_{\m_1\n}
\ee
such that the action (\ref{action}) is invariant
under the gauge transformations (\ref{tj}) with constrained
parameters satisfying (\ref{gfSM}). Note that local Lagrangians for
tensors in irreps of $GL(D,\mathbb R)$ corresponding to Young diagrams
with two rows were studied in operator form by the authors of
\cite{Burdik:2001}.

\section{Arbitrary spin}\label{last}

The geometric equation (\ref{FSbis}) is easily generalized for
symmetric fields $\phi_{\m_1\ldots\m_S}$: when $S$ is {\it odd}
one takes one divergence together with $\frac{S-1}{2}$ traces of
the de Wit-Freedman tensor ${\cal
R}_{\m_1\ldots\m_S\,;\,\n_1\ldots\n_S}$ and when $S$ is {\it
even} one just takes $\frac{S}{2}$ traces \cite{Francia:2002}.
So one constructs a gauge-invariant object with the symmetries of
the spin-$S$ field but containing $2[\frac{S+1}{2}]$ derivatives
($[\frac{S+1}{2}]$ is the integer part of $\frac{S+1}{2}$).
Consequently, Francia and Sagnotti further multiplied by
$\Box^{-[\frac{S-1}{2}]}$ to get a second order field equation.
This natural construction was performed by de Medeiros and Hull
for arbitrary irreducible tensors under $GL(D,\mathbb R)$
\cite{deMedeiros:2002} without dividing by d'Alembertian, i.e.
relaxing the requirement (ii) and preferring locality (i). The
classical field equations in the final de Donder gauge take the
respective forms $\Box \Phi_{\ldots}=0$ and
$\Box^{[\frac{S+1}{2}]} \Phi_{\ldots}=0$. Their set of solutions
is essentially the same since \ba
\ker(\Box)=\ker(\Box^n)\,,\q\forall n\in {\mathbb N}_0\nn\ea in
the space of ``Fourier-transformable" functions, as can be easily
checked in momentum space. In this restricted sense, both
approaches are thus equivalent at the level of sourceless free
field equations. On the one hand the works \cite{Fradkin:1986} of
Fradkin and Vasiliev on higher-spin interactions put some
evidences in favor of the non-local formalism, on the other hand
higher derivative
theories (like Weyl gravity) are known to be subtle to handle at
the quantum level. Nonetheless, duality symmetry could suggest
higher derivatives equations since first Bianchi identities can
become field equations in the dual picture. The possibility of
switching on interactions or quantum consistency might eventually
decide between the physical requirements (i) and (ii).

The geometric equation for a rank $S$ symmetric field is
equivalent to \cite{Francia:2002}\be {\cal
F}_{\m_1\m_2\m_3\m_4\ldots\m_S}=\pr_{(\m_1}\pr_{\m_2}\pr_{\m_3}{\cal
H}_{\m_4\ldots\m_S)}\label{FSgen}\ee which generalizes
(\ref{FS1}). The tensor ${\cal H}_{\m_1\ldots\m_{S-3}}$ is a
non-local function of the field $\phi_{\m_1\ldots\m_S}$ and its
derivatives, whose gauge transformation is proportional to the
trace of the gauge parameter. The gauge-choice ${\cal
H}_{\m_1\ldots\m_{S-3}}=0$ leads to the Fronsdal equation \be
{\cal F}_{\m_1\ldots\m_S}=0\,.\label{Frongen}\ee Basically, the
main supplementary subtlety arising for spin $S\geq 4$ is that the
usual de Donder condition is reachable with a traceless gauge
parameter if and only if the double trace of the field vanishes.
Therefore, in the Fronsdal approach the field is constrained to
have vanishing double trace (which is consistent with the
invariance of the double trace of the field under gauge
transformations with traceless parameter). As pointed out in
\cite{Francia:2002bis}, more work is therefore required to obtain
the double-trace condition for spin $S\geq 4$ in the unconstrained
approach. A solution is to take a modified ({\it identically
traceless}) de Donder gauge which is accessible with constrained
gauge parameters \cite{Francia:2002bis}. After this further
gauge-fixing, the field equation implies the vanishing of the
double trace of the field, thereby recovering the usual de Donder
condition.

It is now straightforward to generalize all our previous results.
The relation (\ref{DD}) is easily generalized by taking $S-2$
curls of the Fronsdal operator ${\cal F}_{\m_1\ldots\m_S}$ to get the
trace of the curvature ${\cal
K}_{\m_1\n_1\,\mid\,\ldots\,\mid\,\m_S\n_S}$ (as is obvious in the
transverse-traceless gauge). Therefore the Fronsdal equation
(\ref{Frongen}) and the Francia-Sagnotti equation (\ref{FSgen})
both imply the tracelessness of the curvature\be \eta^{\n_1\n_2}
{\cal
K}_{\m_1\n_1\,\mid\,\m_2\n_2\,\mid\,\m_3\n_3\,\mid\,\ldots\,\mid\,\m_S\n_S}=0\,.\label{ES}\ee
Conversely, the Einstein-like equation (\ref{ES}) directly implies
the Francia-Sagnotti field equation in its geometric form while,
on the other hand, the ``Poincar\'e lemmas'' of
\cite{Bekaert:2002,Dubois-Violette:1999} allow to derive from
(\ref{ES}) that
\be
{\cal
F}_{\m_1\ldots\m_S}=\pr_{(\m_1}\pr_{\m_2}
\pr_{\m_3}\Sigma_{\m_4\ldots\m_S)}(x)\,.
\label{relation}
\ee
We then recover the Fronsdal approach by an appropriate (partial)
gauge-fixing.

Arbitrary mixed symmetry type tensor gauge fields are studied
similarly, but here the tools developed in \cite{Bekaert:2002}
reveal crucial. The same kind of relationships between the
different approaches is expected to hold. A general recipe for
constructing local Fronsdal-like equations for any exotic tensor
free field is provided by the method followed in section
\ref{exotic}: once the generalized curvature corresponding to a
given exotic gauge field with $S$ columns is given, its trace (on
the first two columns) is identified with $S-2$ curls of the
generalized Fronsdal tensor, the curls being taken on the last
columns. As noticed after (\ref{truc}), an Einstein-like equation
of motion implies \cite{Bekaert:2002} that the corresponding
(irreducible) Fronsdal tensor writes in a way generalizing
(\ref{relation}). Again, an appropriate partial gauge fixing then
brings the Fronsdal tensor to zero.

The duality-symmetric picture sketched in the section
\ref{dualsym} applies for arbitrary mixed symmetry type tensor
fields \cite{Hull:2001,Bekaert:2002} and can be summarized as
follows. For a start, the first Bianchi identities state that the
curvature is irreducible under $GL(D,\mathbb R)$. Then, the
Einstein-like equation means that, on-shell, it is furthermore
irreducible under $SO(D-1,1)$. Next, the crucial mathematical
property is that the Hodge duals of the curvature are therefore
also irreducible under $SO(D-1,1)$. To conclude, the second
Bianchi identities imply that these irreducible tensors are indeed
curvatures for some dual gauge fields.

\section*{Acknowledgements}
%
We thank S. Cnockaert and  M. Tsulaia for interesting discussions.
The work is supported in part by the ``Actions de Recherche
Concert{\'e}es" of the ``Direction de la Recherche Scientifique -
Communaut\'e Francaise de Belgique", IISN-Belgium (convention
4.4505.86), a ``P\^ole d'Attraction Interuniversitaire" (Belgium)
and the European Commission RTN programme HPRN-CT-2000-00131, in
which N.B. is associated with K.U. Leuven.
%


\end{document}